\documentclass{aa}
\usepackage{graphics}
\usepackage{graphics}

\def\spose#1{\hbox to 0pt{#1\hss}}
\def\gsim{\mathrel{\spose{\lower 3pt\hbox{$\mathchar"218$}}
          \raise 2.0pt\hbox{$\mathchar"13E$}}}
\def\lsim{\mathrel{\spose{\lower 3pt\hbox{$\mathchar"218$}}
          \raise 2.0pt\hbox{$\mathchar"13C$}}}

\begin{document}

\title{Is the time lag-luminosity relation of GRBs\\  
a consequence of the Amati relation?}

\author{M. Hafizi\inst{1} and R. Mochkovitch\inst{2} }
\institute{$^1$Tirana University, Faculty of Natural Sciences, Tirana, 
Albania\\
           $^2$Institut d'Astrophysique de Paris - UMR 7095 CNRS 
et Universit\'e Pierre et Marie Curie,\\
98 bis, boulevard Arago, 75014 Paris, France}
\titlerunning{The time lag-luminosity relation of GRBs}

\abstract
{The lag-luminosity relation (LLR)
provides a way of estimating GRB luminosity by measuring the
spectral lags between different energy bands.}
{We want to understand the origin of the LLR
and test its validity. This appears especially important if the
LLR is to be used as a distance indicator.}
{We perform a linear analysis of the lag between two spectral bands.
The lag is obtained as the time interval between the maxima of a 
given pulse in the two bands.}
{We get a simple expression for the lag, which shows in a very simple
way how it is related to the spectral evolution of the burst via the variation
of the peak energy and spectral indices. When this
expression is coupled to the Amati relation, it leads to a LLR that agrees
with the observational results only if the burst's spectral 
evolution is limited to a decrease in peak energy
during pulse decay.
However, when the variation of the spectral indices is also taken into account,
the predicted LLR differs from the observed one.}
{We briefly discuss some ways to solve this problem, such as 
a possible correlation between pulse spikiness and burst luminosity.}

\maketitle

\section{Introduction}
The problem of the distance to gamma-ray bursts remained unsolved until 
the disco\-ve\-ry of the afterglows by Beppo-SAX. Redshifts are now
obtained from optical spectra of the afterglow itself or of the 
host galaxy when the afterglow has faded away. 
Using the  known redshifts, it became 
possible to calibrate relations by linking absolute 
burst outputs (luminosity or total radiated energy) and
quantities directly available from the observations in 
gamma-rays. A Cepheid-like relation bet\-ween variability and luminosity
was proposed, for example, by Reichart et al (2001). More recently,
Atteia (2003) used the Amati relation (Amati et al 2002) to 
introduce ``pseudo-redshifts" which could be useful to rapidly
identify high-$z$ GRBs from their gamma-ray properties alone. In this
paper we concentrate
on the time lag-luminosity relation (LLR) discovered by Norris et al (2000).
The lags were computed by Norris et al using the
burst profiles in BATSE band 1 (20 - 50 keV) and 3 (100 - 300 keV). They
find that the time lag $\Delta t_{13}$ anticorrelates with burst luminosity 
and propose the following power law relation 
\begin{equation}
L=1.3\,10^{53}\ (\Delta t_{13}/0.01\ {\rm s})^{-1.15}\ \ \ {\rm erg.s}^{-1}
\ .
\end{equation}
The origin of the LLR was
then investigated by Kocevski and Liang (2003), Ryde (2005), and 
Ryde et al (2005),
who found that 
the observed lags are a consequence of the burst's spectral evolution.
In this contribution we perform a new analysis of the origin of lags
and discuss how the LLR may be linked to the Amati relation and possibly also 
to the variability-luminosity relation. 
\section{Count rates in different energy bands}
We consider a spectral band $[E_i,\,E_j]$ and assume a spectrum shape
consisting of two smoothly-connected power laws of respective slopes
$\alpha$ and $\beta$ at low and high energy (Band et al, 1993). The count
rate $N_{ij}(t)$ in band $[E_i,\,E_j]$ reads
\begin{equation}
N_{ij}(t)=A(t)\int_{x_i}^{x_j} {\cal B}_{\alpha\beta}(x)\,dx
\end{equation}
where the function $A(t)$ depends on time alone. The li\-mits of the 
integral are $x_{i,j}=E_{i,j}(1+z)/E_{\rm p}(t)$, $z$ being
the redshift of the source, $E_{\rm p}(t)$ the
peak energy of the instantaneous spectrum (in source rest frame)
and ${\cal B}_{\alpha\beta}(x)$
the spectrum shape.

Considering now another spectral band $[E_k,\,E_l]$, we can relate
$N_{kl}(t)$ to $N_{ij}(t)$ in the following way
\begin{eqnarray}
N_{kl}(t) & = & N_{ij}(t)\times {\int_{x_k}^{x_l} {\cal B}_{\alpha\beta}(x)\,dx
\over \int_{x_i}^{x_j} {\cal B}_{\alpha\beta}(x)\,dx}\nonumber\\
& = &
N_{ij}(t)\times {\cal F}_{ijkl}\left[E_{\rm p}(t),\alpha(t),
\beta(t)\right]
\end{eqnarray}
where ${\cal F}_{ijkl}$ can be seen as the ``spectral correction"
between bands $[E_i,\,E_j]$ and $[E_k,\,E_l]$. We simplify the notation
by considering only BATSE bands 1 $[20,50\,{\rm keV}]$ and 3
$[100,300\,{\rm keV}]$ so that we have
\begin{equation}
N_3(t)=N_1(t)\times {\cal F}_{13}\left[E_{\rm p},\alpha,\beta\,\right]
\end{equation}
with
\begin{equation}
{\cal F}_{13}={\int_{100(1+z)/E_{\rm p}}^{300(1+z)/E_{\rm p}} 
{\cal B}_{\alpha\beta}(x)\,dx
\over \int_{20(1+z)/E_{\rm p}}^{50(1+z)/E_{\rm p}} {\cal B}_{\alpha\beta}(x)\,dx}\ .
\end{equation}
We then assume that a given pulse in the burst profile reaches
its maximum at a time $t_1$ (resp. $t_3$) in band 1
(resp. 3) and we estimate the lag by the difference
\begin{equation}
\Delta t_{13}=t_1-t_3\ .
\end{equation}
Since in most cases the observed lags are small compared to the pulse duration, we 
evaluate $\Delta t_{13}$ from a linear analysis of the pulse shape 
around $t_1$. Using ${dN_1\over dt}{\Big |}_{t=t_1}={\dot N}_1(t_1)=0$, we
can write
\begin{equation}
N_1(t)\simeq N_1(t_1)+{1\over 2} {\ddot N}_1(t_1)\,(t-t_1)^2\ ,
\end{equation}
while the spectral correction gives to same order
\begin{equation}
{\cal F}_{13}(t)={\cal F}_{13}(t_1)+{\dot {\cal F}}_{13}(t_1)\,(t-t_1)
+{1\over 2} {\ddot {\cal F}}_{13}(t_1)\,(t-t_1)^2\ ,
\end{equation}
with ${\dot {\cal F}}_{13}(t_1)$ and ${\ddot {\cal F}}_{13}(t_1)$ being related
to the partial derivatives of ${\cal F}_{13}$ with respect to $E_{\rm p}$,
$\alpha$, and $\beta$. For ${\dot {\cal F}}_{13}(t_1)$ we have 
\begin{equation}
{\dot {\cal F}}_{13}(t_1)={\partial {\cal F}_{13}\over \partial E_{\rm p}}
{\Big |}
_{t_1}{\dot E}_{\rm p}(t_1)+
{\partial {\cal F}_{13}\over \partial \alpha}{\Big |}_{t_1}{\dot \alpha}(t_1)+
{\partial {\cal F}_{13}\over \partial \beta}{\Big |}_{t_1}
{\dot \beta}(t_1)\ ,
\end{equation}
while ${\ddot{\cal F}}_{13}(t_1)$ contains nine terms.
We now compute the logarithmic derivative of $N_3(t)$ to the first order in 
$(t-t_1)$
\begin{eqnarray}
& &{{\dot N}_3(t)\over N_3(t)}={{\dot N}_1(t)\over N_1(t)}+
{{\dot {\cal F}}_{13}(t)\over {{\cal F}}_{13}(t)}=
{{\ddot N}_1(t_1)\over N_1(t_1)}(t-t_1)\nonumber \\& + & 
{{\dot {\cal F}}_{13}(t_1)\over {{\cal F}}_{13}(t_1)}
+\left[{{\ddot {\cal F}_{13}(t_1)\over {\cal F}_{13}(t_1)}}
-\left({{\dot {\cal F}_{13}(t_1)\over {\cal F}_{13}(t_1)}}\right)^2\right]
(t-t_1)\ .
\end{eqnarray}
The bracket contains many terms involving partial derivatives of 
${\cal F}_{13}$ to the first
and second order, but it turns out that they are essentially negligible
for the final numerical results. Solving Eq.(10) to get $t_3$, such as
${\dot N}_3(t_3)=0$, finally yields
\begin{equation}
{\Delta t_{13}\over t_{\rm p}}\simeq {f_{13,E}\,{\dot e}_{\rm p}+
f_{13,\alpha}\,{\dot a}+f_{13,\beta}\,{\dot b}\over C_1}
\end{equation}
with
\begin{eqnarray}
& &f_{13,X}={\partial Log {\cal F}_{13}\over \partial Log X}{\Big |}_{t_1},\ \ 
{\dot e}_{\rm p}={{\dot E_{\rm p}}\over E_{\rm p}}\,t_{\rm p}\nonumber\\ 
& &{\dot a}={{\dot \alpha}\over \alpha}\,t_{\rm p},\ \ 
{\dot b}={{\dot \beta}\over \beta}\,t_{\rm p}\ \ 
{\rm and}\ \ 
{C_1\over t_{\rm p}^2}={{\ddot N}_1(t_1)\over N_1(t_1)}
\end{eqnarray}
where $t_{\rm p}$ is the characteristic duration of the pulse.
For two given spectral bands and an assumed spectral shape, Eq.(11) provides 
a linear estimate of the lag, which directly shows how it is related to
burst spectral evolution via the temporal
derivatives of $E_{\rm p}$, $\alpha$ and $\beta$. The ``curvature parameter"
$|C_1|$ depends on the pulse shape at maximum, large (resp. small) $|C_1|$
values corresponding to spiky (resp. broad) pulses.
\section{The lag luminosity relation}
Equation (11) gives the lag between BATSE bands 1 and 3 if the values of $E_{\rm p}$,
$\alpha$, $\beta$, their time derivatives, and the pulse shape are known
at maximum. It will become a LLR if these parameters
can be related in some way to the luminosity. The Amati relation 
(Amati et al, 2002) provides
such a link but, in its most studied version, it connects the isotropic 
energy in gamma-rays to the 
$E_{\rm p}$ value of the global spectrum. However, it has been suggested 
that a similar relation may exist between $E_{\rm p}$ and the luminosity. 
Yonetoku et al (2004) find, 
for example, a relation between the maximum luminosity 
and the global $E_{\rm p}$,
while Ghirlanda et al (2005) propose a relation between the 
values 
of $E_{\rm p}$
and the luminosity both taken at pulse maximum
\begin{equation}
E_{\rm p}=380\left({L\over 1.6\,10^{52}\ {\rm erg.s}^{-1}}\right)^{0.43}
\ {\rm keV}\ . 
\end{equation}
If a substantial fraction of bursts satisfy Eq.(13), 
it will, together with Eq.(11), lead to a LLR that 
can be compared to the observational data.
\subsection{A (too) simple example}
\begin{figure*}{}
\begin{center}
\begin{tabular}{cc}
\resizebox{0.58\hsize}{!}{\includegraphics{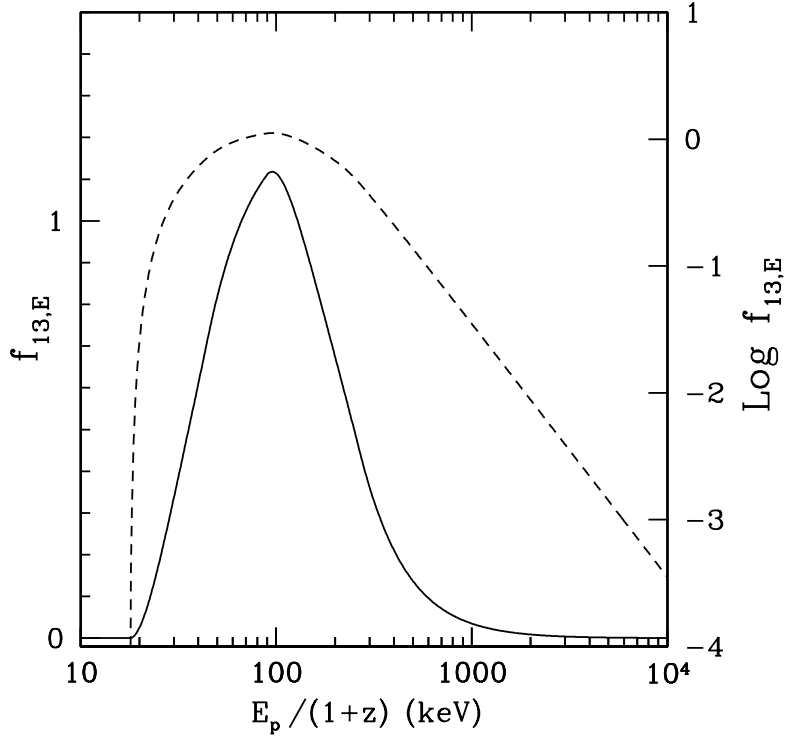}}
\end{tabular}
\end{center}
\caption{The function $f_{13,E}$ given by Eq.(14) plotted as a function of the 
observed peak energy for $\alpha=-1$ and $\beta=-2.25$; full line:
linear scale; dashed line: logarithmic scale.}
\end{figure*}
\begin{figure*}{}
\begin{center}
\begin{tabular}{cc}
\resizebox{0.58\hsize}{!}{\includegraphics{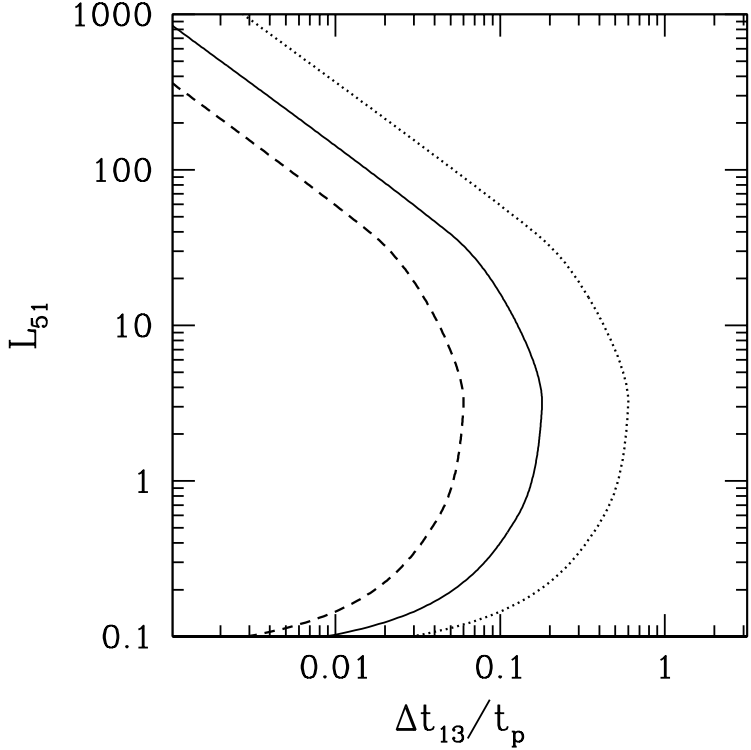}}
\end{tabular}
\end{center}
\caption{Lag-luminosity relation resulting from Eq.(14) which limits
the spectral evolution
to a decrease of $E_{\rm p}$ during pulse decay
($\alpha$ and $\beta$ being kept constant 
respectively equal to $-1$ and $-2.25$). 
The three lines cover an order of magnitude in 
$|{\dot e}_{\rm p}/C_1|$ from 0.1 (dashed line) to 0.3 (full line)
and 1 (dotted line). A redshift value $z=1$ has been assumed. }
\end{figure*}
In this section we limit the burst's spectral evolution to the variation of
$E_{\rm p}$ only and write ${\dot a}={\dot b}=0$. Equation (11) then simply becomes
\begin{equation}
{\Delta t_{13}\over t_{\rm p}}\simeq {f_{13,E}\,{\dot e}_{\rm p}
\over C_1}\ \ \ \ {\rm with}\ \ \ \  f_{13,E}=
{\partial Log {\cal F}_{13}\over 
\partial Log E_{\rm p}}{\Big |}_{t_1,\alpha,\beta}\ .
\end{equation} 
The function $f_{13,E}$ tends towards 0 for the 
highest and lowest values of $E_{\rm p}$, while
it is at its maximum for $50\lsim E_{\rm p}/(1+z)\lsim 100$ keV.
This is illustrated in Fig.1 and 
can be easily understood, since for $E\ll E_{\rm p}$ 
(resp. $E\gg E_{\rm p}$) the 
Band function ${\cal B}(x)$ behaves as $x^{\alpha}$ (resp. $x^{\beta}$).
Therefore for high $E_{\rm p}$ values ($E_{\rm p}/(1+z)\gg 300$ keV)
\begin{equation}
{\cal F}_{13}\simeq {300^{1+\alpha}-100^{1+\alpha}
\over 50^{1+\alpha}-20^{1+\alpha}}=C^{st}\ \ \ {\rm and}\ \ \ f_{13,E}=0
\end{equation} 
a similar result with $\beta$ replacing $\alpha$ being obtained for low 
$E_{\rm p}/(1+z)\ll 20$ keV. At intermediate values, 
$50\lsim E_{\rm p}/(1+z)\lsim 100$ keV,
and making the rough approximation that ${\cal B}(x)\propto x^{\alpha}$ in
band 1
while ${\cal B}(x)\propto x^{\beta}$ in band 3, we obtain
\begin{equation}
{\cal F}_{13}\simeq {300^{1+\beta}-100^{1+\beta}
\over 50^{1+\alpha}-20^{1+\alpha}}\left({E_{\rm p}\over 1+z}\right)
^{\alpha-\beta}
\end{equation}
and therefore
\begin{equation}
f_{13,E}=\alpha-\beta\ .
\end{equation}
Finally it can be shown that the shape of ${\cal B}(x)$ at small $x$, ${\cal B}(x)
=x^{\alpha}\left[1-(2+\alpha)x\right]$ 
leads to a 
power law behavior for $f_{13,E}$ at large $E_{\rm p}$ as seen in Fig.1.

If the Amati-like relation  
(Eq.(13)) is satisfied, high luminosity bursts will have a large
$E_{\rm p}$ and therefore a small lag, while lags will be comparatively large
for bursts with $E_{\rm p}$ in the range 50\,(1+z) -- 100\,(1+z) keV,
i.e. $L\sim$ $4\,10^{50}\,(1+z)^{2.3}$ erg.s $^{-1}$ at pulse maximum. 
Equation (14) also predicts that, for a given 
luminosity and spectral evolution, spiky bursts (large $|C_1|$) will 
have smaller lags than bursts with broad pulses (small $|C_1|$), in
agreement with observations
(Hakkila \& Giblin 2006). 
The LLR obtained with this simple
model is represented in Fig.2 
for different values of the ratio $|{\dot e}_{\rm p}/C_1|$ and 
a typical burst redshift $z=1$. A large (resp. small) ratio 
corresponds to a faster (resp. slower) spectral evolution or to
a broader (resp. spikier) pulse. 
At high luminosity ($L> 10^{52}$ erg.s$^{-1}$), the LLR has a power law
behavior since
\begin{equation}
{d\,Log L\over d\,Log \Delta t}=\left(
{d\,Log L\over d\,Log E_{\rm p}}\right)\left(
{d\,Log E_{\rm p} \over d\,Log  \Delta t}\right)
\end{equation}
the first factor being given by the Amati relation (Eq.13), 
while the second
results from the spectrum shape. From Eq.(14) we have
\begin{equation} 
{d\,Log E_{\rm p} \over d\,Log  \Delta t}=
{d\,Log E_{\rm p} \over d\,Log f_{13,E}}
\end{equation}
which is a constant at high $E_{\rm p}$ (and hence luminosity) 
values (see Fig.1). However at luminosities smaller than $10^{52}$ erg.s$^{-1}$ 
the model predicts that the power law behavior of the LLR
should break down with the lag passing through a maximum and then 
decreasing. This clearly contradicts GRB 980425, which 
has both a very low luminosity and a large lag. But GRB 980425 does not
satisfy the Amati relation so that its departure from the LLR is not
surprising. One should instead consider that this burst has a broad temporal
profile, i.e. a small $|C_1|$ and an $E_{\rm p}$ of $138$ keV
(Ghisellini et al 2006)
corres\-pon\-ding to the maximum of $f_{13,E}$ (see Fig.1) and 
therefore to a large expected lag. 
\subsection{A more complete study}
The spectral evolution of GRBs is, however, not limited to a decrease in 
$E_{\rm p}$ during pulse decay. A hard-to-soft evolution is 
also observed for the spectral indices $\alpha$ and $\beta$. 
In some extreme cases $\alpha$ has been seen to decrease from $\alpha\sim 1.5$
(a value a priori excluded by the synchrotron model) to about $-0.5$
in just a few seconds (Crider et al 1997). 
When 
the variation in the spectral indices is included in our linear analysis, 
it no longer predicts a vanishing lag for high or low
$E_{\rm p}$ and $L$ values since now $f_{13,\alpha}\ne 0$ when
$E_{\rm p}\rightarrow \infty$ and $f_{13,\beta}\ne 0$ when
$E_{\rm p}\rightarrow 0$ 
(for example $\lim_{E_{\rm p}\to \infty}
f_{13,\alpha}\simeq -1.7$ for 
$\alpha=-1$ and $\beta=-2.25$).
The lag then reaches a constant
limiting value at low and high luminosities, where it apparently
contradicts the observed LLR. This is shown in Fig.3 where
our calculated LLR has been plotted for different values of $\dot a$
and a fixed $\dot b=0.1$.
Even a moderate variation in the spectral indices has 
a dramatic effect on the 
LLR and the global agreement with the Norris et al (2000) results that
was found in the last section is now lost. 
\begin{figure*}{}
\begin{center}
\begin{tabular}{cc}
\resizebox{.6\hsize}{!}{\includegraphics{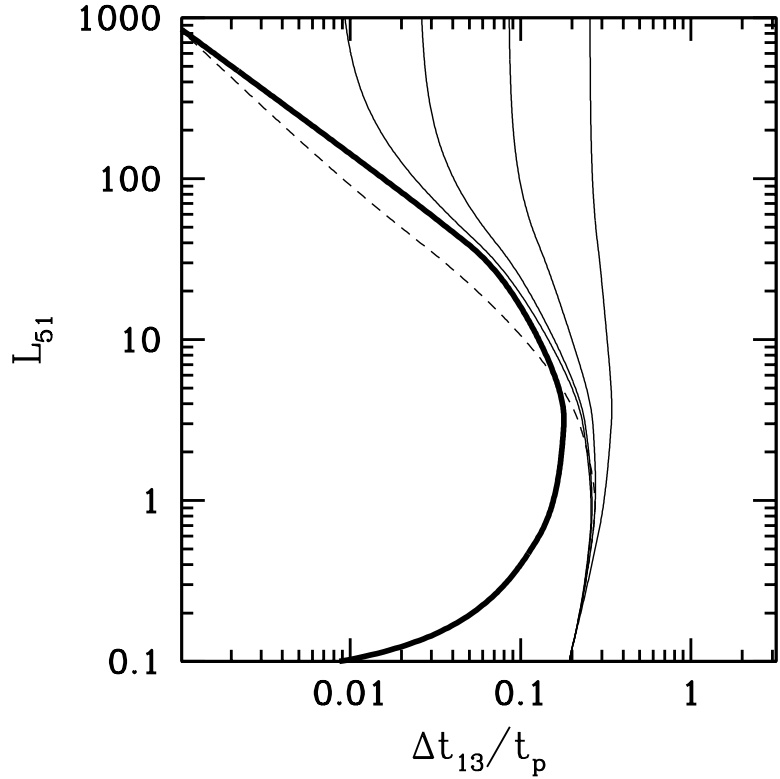}}
\end{tabular}
\end{center}
\caption{LLR with $|{\dot e}_{\rm p}/C_1|=0.3$, but now also
including the variation in the 
low and high-energy spectral indices. The thin lines 
correspond respectively (from left to right) to  
$\dot a=0.01$, 0.03, 0.1, and 0.3 (and have all $\dot b=0.1$), while
the thick full line represents the $\dot a=\dot b=0$ case. 
The dashed line is obtained with $\dot a=\dot b=0.1$ and a varying
curvature parameter given by Eq.(20).}
\end{figure*} 
\section{Discussion }
These results clearly disagree with the 
observational data for the most luminous GRBs. Therefore if real bursts
do satisfy the LLR proposed by Norris et al. (2000), 
a solution has to be found for the apparent discrepancy 
between our analysis  and the observations:
\vskip 0.3cm
(i) A first option could be that in most cases the variation in the spectral index
$\alpha$ is small, at least around pulse maximum. However this does not
seem to be the case for the bright events for which a detailed, time 
resolved, spectral analysis has been possible (Preece et al 2000). 
Moreover, the constraint on any variation in $\alpha$ appears so severe
(only the LLR with ${\dot a}=0.01$ in Fig.3 is marginally compatible with
the Norris et al. results) that it seems difficult to expect it will be
satisfied by a large fraction of GRBs.
\vskip 0.3cm
(ii) A more interesting possibility would be that a relation may exist
between the curvature parameter $|C_1|$ and the luminosity, bursts
with spiky pulses being more luminous than bursts with broad pulses.
This might be a different way to express the variability-luminosity
relation proposed by Reichart et al (2001). We tried, for example,
a simple linear expression of the form
\begin{equation}
|C_1|=2+0.2 (L_{51}-1)
\end{equation}
where the resulting LLR with $|{\dot e}_{\rm p}/C_1|=0.3$ and ${\dot a}=
\dot b=0.1$
is represented in Fig.3. In spite of the variation in $\alpha$,
it now gives again very small lags at high luminosity because the pulses
are then much spikier than at low luminosity.
\section{Conclusion}
We have performed a linear analysis of the time lag between two 
spectral
bands and have obtained a simple relation (Eq.(11)) which clarifies 
how the lag is related to the burst spectral
evolution. When this relation is used in conjunction with the Amati relation
it leads to a satisfactory LLR only if the spectral evolution of GRBs is
limited to a decrease of $E_{\rm p}$ during pulse decay.
If the variation in the spectral indices
is also included, the lag does not decrease any longer to low values,
even at very high burst
lumi\-no\-si\-ty. We have briefly discussed the possibility that short 
lags might be
recovered if burst luminosity is correlated to the shape of the pulses, 
bursts with spiky pulses being more luminous than bursts with broad pulses.
The available sample of GRBs with both measured lags and known distance
is still small but should increase with SWIFT. This will allow constraining
tests of the results presented in this paper.

\begin{acknowledgements}
The authors thank Fr\'ed\'eric Daigne for helpful discussions and 
Felix Ryde for 
communicating several of his results prior to publication. 
They also thank the anonymous referee whose comments have considerably
improved an initial version of this paper.
\end{acknowledgements}

\end{document}